\documentstyle[prl,aps,multicol]{revtex}
\input{epsf}
\begin{document}
\draft
\title{Quantum localization in rough billiards}

\author{Klaus M. Frahm and Dima L. Shepelyansky$^{*}$}

\address {Laboratoire de Physique Quantique, UMR C5626 du CNRS, 
Universit\'e Paul Sabatier, F-31062 Toulouse Cedex 4, France}

\date{28 November 1996}

\maketitle

\begin{abstract}
We study the level spacing statistics $p(s)$ and eigenfunction properties 
in a billiard with a rough boundary. Quantum effects lead to 
localization of classical diffusion in the angular momentum space and 
the Shnirelman peak in $p(s)$ at small $s$. The ergodic regime with 
Wigner-Dyson statistics is identified as a function of 
roughness. Applications for the $Q$-spoiling in optical resonators are 
also discussed. 
\end{abstract}


\begin{multicols}{2}
\narrowtext


In 1984, Bohigas, Giannoni, and Schmit \cite{bohigas} demonstrated 
that random matrix theory (RMT) \cite{mehta} describes the level spacing 
statistics of classically chaotic billiards. After that such types of 
billiards have been studied in great detail by different groups 
\cite{leshouches}. However, all billiards under investigation 
were characterized only by one typical time scale $t_c$, namely, the time between 
collisions with the boundary. Another type of chaotic systems with diffusive 
behavior, like the kicked rotator \cite{rotator}, has an additional much longer 
time scale $t_D \gg t_c$ which is required to cover the accessible 
phase space. In this situation quantum interference effects may lead 
to exponential dynamical localization of the eigenstates 
and disappearance of level repulsion. 

Recently, it has been shown \cite{fausto} that the diffusive regime also 
appears in a nearly circular Bunimovich stadium billiard. 
The authors of \cite{fausto} gave an estimate for the localization 
length in the angular momentum space and found the energy border 
$E_{erg}$ above which the eigenstates become ergodic on the energy surface. 
Their numerical results demonstrate the change of level statistics 
from Wigner-Dyson to Poisson when the energy becomes smaller than 
$E_{erg}$. However, this example being very interesting for mathematical 
studies is not physically realistic. 

At the same time a great progress has been reached in optics of 
microcavities like micrometer-size droplets \cite{droplet} and microdisk 
lasers \cite{microdisk}. The industrial request to produce 
directed light pushed 
the researches to investigate ray dynamics in weakly deformed circular 
billiards and droplets. It was shown that above some critical 
deformation the ray dynamics becomes chaotic. As a consequence the 
directionality of light from droplets and $Q$-factors of such resonators 
are strongly affected \cite{stone}. However, due to smoothness of the 
deformation the diffusive 
regime was hardly accessible in such systems. 

In this Letter, we investigate another type of generic boundary deformation 
which may have important physical applications. Namely, we consider 
billiards with a rough boundary. In this approach, the boundary is a 
random surface with some finite correlation length. The physical realizations 
of such situation can be quite different. As examples, we can mention 
surface waves in the droplets which are practically static for the light 
\cite{droplet}, nonideal surfaces in microdisk lasers \cite{microdisk}, 
and capillary waves on a surface of small metallic clusters \cite{clusters}. 
On a first glance it seems that such a rough boundary in a circular 
billiard will destroy the conservation of angular momentum leading to 
ergodic eigenstates and RMT level statistics. In spite of this we show 
that there is a region of roughness in which the classical dynamics is 
chaotic but the eigenstates are localized and the level spacing statistics 
$p(s)$ has the sharp Shnirelman peak at small spacings $s$ 
\cite{shnirel,chirikov1}. We also demonstrate the close relation between this 
model and the kicked rotator. 


As a model of a rough billiard we chose a circle with a deformed elastic 
boundary given by $R(\theta)=R_0+\Delta R(\theta)$ with $\Delta R(\theta)/R_0=
\mbox{Re}\ \sum_{m=2}^M \gamma_m\,e^{i m \theta}$ and where $\gamma_m$ 
are random complex coefficients. The roughness of the surface is 
given by $\kappa(\theta)=(d R/d\theta)/R_0$. In the following, we will 
consider the case of weak roughness $\kappa \ll 1$. Furthermore, one can 
model different type of surfaces by chosing the appropriate dependence of 
the amplitudes on $m$. However, our results show that in the domain of 
strong chaos the classical diffusion and quantum localization in 
orbital momentum space are determined by the 
angle average $\tilde\kappa^2=\langle \kappa^2(\theta)\rangle_\theta$. 
Due to that 
we choose a typical dependence $\gamma_m \sim 1/m$ such that all harmonics 
give the same contribution in the roughness. 

\begin{figure}
\epsfxsize=3in
\epsfysize=2in
\epsffile{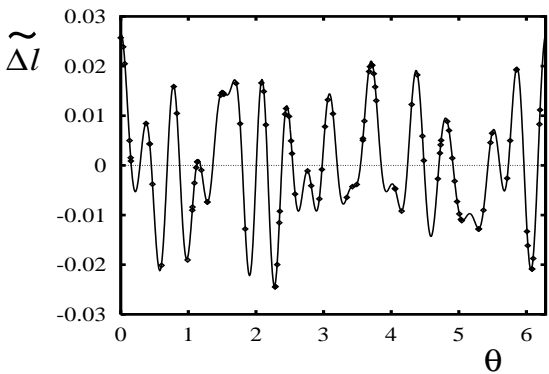}
\vglue 0.2cm
\caption{Comparison of the exact dynamics (points) and the rough map 
(\protect\ref{map}) (full curve $\kappa(\theta)$). The points correspond 
to $\tilde{\Delta l}=(\bar l -l)/(2\sqrt{l_{max}^2-l^2})$ and $M=20$, 
$\tilde\kappa=0.011$.} 
\label{fig1}
\end{figure}

First, we consider the classical ray dynamics which for $\kappa\ll 1$ can be 
described by the following rough map:
\begin{eqnarray}
\label{map}
\bar l & = & l + 2 \sqrt{l_{max}^2-l_r^2}\ \kappa(\theta)\ ,\cr
\bar \theta & = & \theta+\pi-2\,\mbox{arcsin}(\bar l/l_{max})
\ \phantom{\bigg(}.
\end{eqnarray}
Here the first equation gives the change of the angular momentum (and 
velocity vector) due to the collision with the boundary and the second one 
the change of angle between the collisions. This map describes  
the dynamics in the vicinity of the resonant value $l_r$ defined by 
the condition $\bar\theta=\theta+2\pi r$ with integer $r$.  Our numerical 
simulations of the exact ray dynamics show that the rough map (\ref{map}) 
indeed gives an excellent description (see Fig. \ref{fig1}). A similar 
map for a stadium billiard was given in \cite{fausto}. However, in contrast 
to \cite{fausto}, where $\kappa(\theta)$ has a discontinuity, the global chaos 
sets in only if the roughness is above 
some critical value $\sqrt{M} \Delta R/R\sim \tilde\kappa > \kappa_c$. Below $\kappa_c$ the 
KAM theory is valid and the phase space is divided by invariant curves. 
The chaos border can be estimated on the basis of Chirikov criteria 
of overlapping resonances \cite{chirikov2} which gives $\kappa_c\sim 
4 M^{-5/2}$ (the numerical coefficient is extracted from the data 
for $M=20$). This border drops strongly with $M$ and 
therefore we will concentrate on the regime of strong chaos without 
visible islands of stability (case of Fig. \ref{fig1}). In this 
regime the spreading in angular momentum space goes in a diffusive way 
with a diffusion constant that is easily estimated from (\ref{map}) as 
$D=(\Delta l)^2/\Delta t=4\,(l_{max}^2-l_r^2)\,\tilde \kappa^2$ where 
the time $t$ is measured in number of collisions. 

\begin{figure}
\epsfxsize=3in
\epsfysize=2.5in
\epsffile{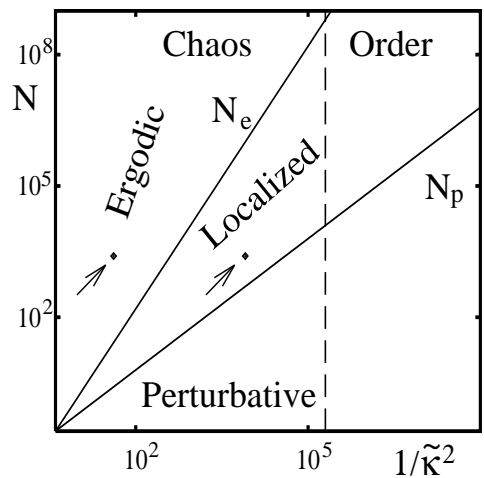}
\vglue 0.2cm
\caption{Global diagram showing the eigenstate properties for different 
values of level number $N$ and roughness $\tilde\kappa$ for $D\gg M$. 
The two full 
lines give the ergodic ($N_e$) and perturbative ($N_p$) borders, the dashed 
line is the classical chaos border $\kappa_c\approx 0.002$ for $M=20$. 
The parameters of Figs. \protect\ref{fig3},\protect\ref{fig4} are 
shown by the two points with arrows.}
\label{fig2}
\end{figure}

Now, we turn to the investigation of the quantum problem. For this we 
expand the wave function $\psi(r,\theta)$ with energy $E=\hbar^2 k^2/2m$ 
in terms of the Hankel functions which form a complete set:
\begin{equation}
\label{hankel}
\psi(r,\theta)=\sum_l \left(a_l\,H_{|l|}^{(+)}(kr)\,e^{il\theta} +
b_l\,H_{|l|}^{(-)}(kr)\,e^{il\theta} \right)\ .
\end{equation}
The regularity of $\psi$ at $r=0$ requires $a_l=b_l$ so that the $a_l$ 
are the amplitudes in angular momentum space. The boundary condition 
$\psi(R(\theta),\theta)=0$ results in a second equation $b_l=\sum_{l'} 
S_{ll'}(E)\,a_{l'}$ where $S_{ll'}(E)$ has the meaning of a scattering 
matrix for waves reflected at the rough boundary. The energy eigenvalues 
are determined by $\det[1-S(E)]=0$. A convenient expression for the 
S-matrix can be obtained by the quasiclassical approximation 
$H_l^{(\pm)}(kr)\approx 2\,[2\pi k_l(r) r]^{-1/2}\ \exp[\pm i 
(\mu_l(r)-\pi/4)]$ 
where $k_l(r)=k (1-r_l^2/r^2)^{1/2}$, $\mu_l(r)=\int_{r_l}^r\,d\tilde r\,
k_l(\tilde r)$ and $r_l\approx |l|/k$ is the classical turning point. From 
this representation and the boundary condition for $\psi$ we obtain
(for more details see \cite{frahm}):
\begin{equation}
\label{smatrix}
S_{ll'}\approx e^{i\mu_l(R_0)+i\mu_{l'}(R_0)-i\pi/2}\ 
<l| e^{i\,2k_{l_r}(R_0)\,\Delta R(\theta)}|l'>\ .
\end{equation}
This is a local unitary expression for $S$ near $l_r$ which is valid for 
$\Delta l=|l-l'|\ll l_{max}=k R_0$ and $M< l_{max}$. A stationary phase 
approximation for the 
$\theta$-integral gives the classical change of $l$ [see (\ref{map})] and 
determines the structure of $S$-matrix. 
In fact, this matrix is very similar to the evolution operator of the 
kicked rotator \cite{rotator} corresponding to $\Delta R\propto \cos\theta$. 
According to this analogy the localization length $\ell$ is determined 
by the classical diffusion rate $\ell=\beta D/2$ 
where $\beta$ is the symmetry index for orthogonal ($\beta=1$) \cite{shep}
or local unitary symmetry ($\beta=2$) \cite{bluemel}. Since generally 
$\Delta R(\theta)\neq \Delta R(-\theta)$, we have $\beta=2$ so that 
the localization length is directly determined by the roughness

\begin{equation}
\label{loc_len}
\ell = D = 4\,(l_{max}^2-l_r^2)\,\tilde\kappa^2. 
\end{equation}
This result can also be derived on a more rigorous ground based on the 
supersymmetry approach for a model with random phases $\mu_l$ \cite{frahm}. 
The expression (\ref{loc_len}) is only valid for $D > M$ 
while $1<D<M$ corresponds to a more complicated regime 
with $\ell \sim M$ (see below).
For $1 < D \ll l_{max}$ the eigenfunctions are exponentially localized in the 
orbital space $|a_l|\propto \exp(-|l-l_0|/\ell)$ while the classical 
dynamics is ergodic on the whole energy surface.  The quantum dynamics becomes ergodic only 
for $\ell>l_{max}$. According to the Weyl formula the level number at 
energy $E$ is $N\approx m R_0^2 E/2\hbar^2=l_{max}^2/4$. Therefore the states 
are ergodic for 
\begin{equation}
\label{n_erg}
N>N_e\approx \frac{1}{64\,\tilde \kappa^4}\ .
\end{equation}
This border is much higher than the perturbative border $N<N_p\approx 
1/(16 \tilde\kappa^2)$ where the diffusion mixes less than one state 
($D\approx 1$). These different 
regimes are presented in Fig. \ref{fig2}. 

\begin{figure}
\epsfxsize=3in
\epsfysize=2in
\epsffile{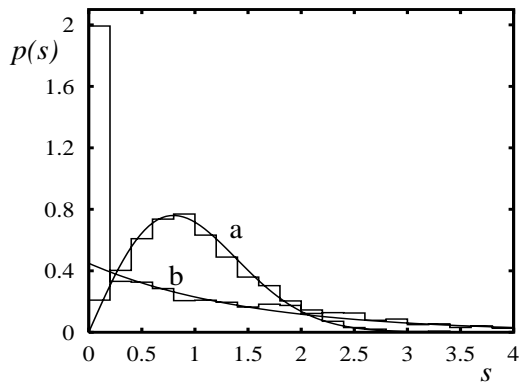}
\vglue 0.2cm
\caption{Level spacing distribution $p(s)$ for $l_{max}\approx 100$, 
$M=20$ and average roughness: a) $\tilde\kappa\approx 0.16$ (ergodic, 
$\ell\approx 1000$) and b) $\tilde\kappa\approx 0.012$ (localized, 
$\ell\approx 6$). The total statistics is 2000 levels 
($2150<N<2350$) from 10 different realizations of the rough boundary. 
Also shown are the Wigner-Dyson and rescaled Poisson distributions with 
$\alpha=0.33$.}
\label{fig3}
\end{figure}

\begin{figure}
\epsfxsize=3in
\epsfysize=2.2in
\epsffile{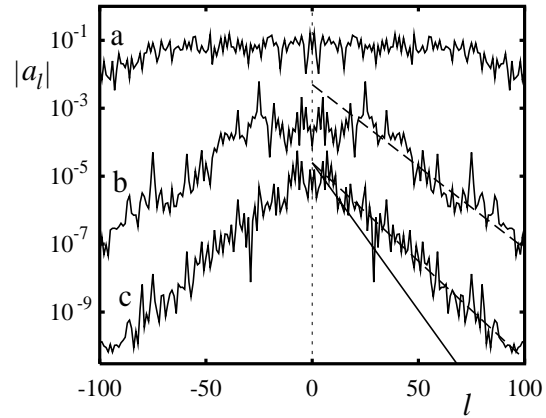}
\vglue 0.2cm
\caption{The amplitudes $|a_l|$ for three typical eigenfunctions 
corresponding to the parameters of Fig. \protect\ref{fig3} 
(with level number $N\sim 2250$):
a) ergodic case; b) localized quasidegenerate state with $s=0.1$ and 
fitted localization length $\ell\approx 9$ (dashed line); 
c) nondegenerate localized state with fitted $\ell\approx 7.5$ 
(dashed line). The full line shows the theoretical length 
[see Eq. (\protect\ref{loc_len})] 
$\ell\approx 5$. The rough boundary realization 
is the same as in Fig. \protect\ref{fig1} (b,c) or appropriately rescaled 
(a). The curves b and c are shifted by factors $10^{-2}$ or $10^{-4}$, 
respectively. }
\label{fig4}
\end{figure}

To check the above theoretical predictions, we have solved numerically 
the boundary condition for $\psi(r,\theta)$ (\ref{hankel}). In this way, 
we have obtained both the energy eigenvalues and the amplitudes $a_l$ 
(with normalization $\sum_l |a_l|^2=1$). 
In the ergodic regime $N>N_e$, we find that the level spacing statistics 
$p(s)$ is in a good agreement with RMT (see Fig. \ref{fig3}). On the contrary, 
in the localized regime $N<N_e$, approximately each second level 
is quasidegenerate leading to the Shnirelman peak \cite{shnirel} at 
small spacings (Fig. \ref{fig3}). This peak represents approximately 
a fraction $\alpha\approx 0.33$ of all spacings. The other spacings are 
described by a rescaled Poisson distribution $p(s)=(1-\alpha)^2\,
\exp[-(1-\alpha)\,s]$. The appearance of the Shnirelman peak is in agreement 
with the prediction made in \cite{chirikov1}. Its physical origin is the 
time reversal symmetry ($S_{-l,-l'}=S_{ll'}^*$ or $a_{-l}=a_l^*$) 
due to which two states localized around $l_0$ 
and $-l_0$ form a quasidegenerate pair of symmetric and antisymmetric 
states \cite{footnote}. The original Shnirelman theorem was formulated for quasi-integrable 
billiards \cite{shnirel}. Our case corresponds to the chaotic domain, however, 
due to localization the peak still exists. A similar situation for the 
kicked rotator was studied in \cite{chirikov1}. The fraction of non degenerate 
levels ($1-2\alpha\approx 0.34$) is due to states localized near 
$l_0\approx 0$. The measure of such states is approximately 
$4\ell/l_{max}\approx 1-2\alpha$ where the numerical coefficient was 
extracted from our data (see Fig. \ref{fig4}). This peak was not found 
in \cite{fausto} because the stadium billiard has additional symmetries 
and only the states of one parity were considered. 

In Fig. \ref{fig4}, we show three typical eigenfunctions in 
angular momentum representation. In the ergodic case, the probability is 
homogeneously distributed nearly in the whole interval $(-l_{max},l_{max})$. 
We mention that in our computations we have included about $10$ evanescent 
modes (with $|l|>k R_0$). The two other wave functions are in the 
localized regime of Fig. \ref{fig2} with $\ell \ll l_{max}$. One of 
them has a double peak structure due to tunneling between 
time reversed angular momentum states. This state corresponds to a 
quasidegenerate Snirelman state. We also observed many other similar 
states with much smaller level splitting ($s<10^{-2}$). The other 
localized state has the maximum $l_0\approx 0$ and corresponds to 
a nondegenerate level. Both states clearly show exponential 
localization indicated by the dashed lines. The numerical values 
of the localization length are a bit higher than the 
estimate (\ref{loc_len}) with $l_r\approx l_0\ll l_{max}$. 

\begin{figure}
\epsfxsize=3in
\epsfysize=2in
\epsffile{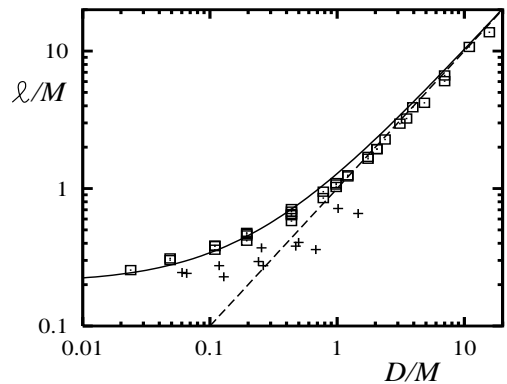}
\vglue 0.2cm
\caption{Rescaled localization length $\ell/M$ as a function of 
rescaled diffusion rate $D/M$ for $l_{max}=100$: crosses 
are numerical data for 
billiards with $0.008< \tilde\kappa<0.027$ and $20\le M \le 60$; 
squares are data for the effective kicked rotator model (see text) with 
$0.006< \tilde\kappa<0.043$, $5\le M \le 60$ and a matrix size of $600$; 
the full curve 
is the fit $\ell=0.5 M /\ln[1+0.5 /(0.05+D/M)]$ with the asymptotic limit 
$\ell=D$ shown by the dashed line.}
\label{fig5}
\end{figure}

To understand this discrepancy, we have calculated $\ell$ for 
a wide region of $M$ and $\tilde\kappa$. Many cases were systematically 
different from (\ref{loc_len}). To increase the parameter range 
we also studied the effective kicked rotator model with 
random phases $\mu_l$ in the evolution operator (\ref{smatrix}) 
(and $l_{max}=100$, $l_r=0$). The results of these studies demonstrate 
the scaling behavior $\ell/M=f(D/M)$ (see Fig. \ref{fig5}). We attribute the 
remaining difference between the two models to a smaller sample size 
for the billiard ($100$ vs. $600$) and finite $l_r\sim l_{max}$ values
there. 
For $D\gg M$ the scaling reproduces the estimate (\ref{loc_len})  
while for $D <M$ the length $\ell$ remains close to $M$. In this case, 
the evolution operator becomes a band random matrix of width $M$ with 
strong {\it diagonal} fluctuations. They are much larger than the off-diagonal 
matrix elements so that $\ell\sim M$ for $1<D<M$ \cite{brm}. 


Let us now discuss how the above properties of eigenstates will 
affect the characteristics of resonators with rough 
boundaries. In optical resonators the rays with large reflection 
angle escape from the system (see for example \cite{stone}). Due to 
that there is an effective absorption in the momentum space for 
$l<l_c$ where the critical momentum is determined by 
the index of reflection so that typically $l_c/l_{max}\approx 1/2$. 
This absorption affects the $Q$-value of the resonator which is 
approximately equal to the number of collisions until the escape. In 
the ergodic regime this number is determined by diffusive spreading in 
the momentum space so that $Q\sim l_c^2/D \sim \tilde\kappa^{-2}$ 
being proportional to the inverse Thouless energy. 
On the other hand in the localized regime the probability to reach 
$l_c$ is exponentially suppressed. The subsequent estimate gives 
$\ln Q\sim l_c/\ell \sim 1/(l_{max}\,\tilde\kappa^2) > 1$ and determines 
the roughness border $\tilde\kappa_Q$ in the $Q$-spoiling which corresponds 
to our ergodic border (\ref{n_erg}). For given $l_{max}$ this border 
is $\tilde\kappa_Q\approx 1/(2\,\sqrt{l_{max}})$. 

Above we discussed the effects of the rough boundary in a circular billiard. 
However, we can argue that similar effects should be observed in more general 
types of convex smooth billiards. Indeed, in such systems, a large 
fraction of the phase space is integrable and characterized by two 
quantum numbers. As in the circular case the rough boundary 
will lead to a diffusive behavior in one of the quantum numbers parallel to 
the energy surface and again the quantum interference effects can give its
localization. We note 
that our results are valid for weakly rough billiards $\kappa\ll 1$ 
while the case of strong roughness $\kappa\sim 1$ deserves separate 
studies. 

A generalization to the three dimensional (3d) case represents an interesting 
direction for further research. We expect that for a sphere 
with rough boundary the above 2d analysis is very relevant. Indeed, 
in a perfect sphere 
a trajectory is confined to a plane and the precession frequency of 
this plane is zero. Due to roughness this plane will slowly 
precess with a frequency proportional to $\kappa$. This adiabatic 
process will weakly affect the dynamics and quantum localization 
inside the plane section. 
The localization should give rise to the Shnirelman peak in 3d.
However, the situation should be quite different in more general 3d smooth 
integrable billiards (e.g.~ellipsoid). There the precession frequency is 
rather high and the diffusion becomes really two dimensional on the 
energy surface. As for localization in 2d disordered systems, 
one might expect an exponential fast increase of $\ell$ with 
the roughness ($\ln \ell\sim l_{max}^2\,\kappa^2$). 

In conclusion, we studied the billiards with strongly different diffusive 
and instability time scales. In this situation quantum effects can produce 
localized eigenstates breaking classical ergodicity on the energy surface. 
We wonder if the periodic orbit approach can lead to a deeper understanding 
of the physical situation.

\end{multicols}

\end{document}